\documentclass[aps,twocolumn,prl,amsmath,amssymb,showpacs,notitlepage,superscriptaddress]{revtex4-1}
\usepackage{graphicx}
\usepackage{dcolumn}
\usepackage{bm}
\usepackage{wrapfig}
\usepackage{amsmath,amsfonts,amssymb,bm,ulem}
\usepackage[usenames]{color}

\begin{document}

\title{Tunable sign change of spin Hall magnetoresistance in Pt/NiO/YIG structures}

%



\author{Dazhi Hou}
\affiliation{WPI Advanced Institute for Materials Research, Tohoku University, Sendai 980-8577, Japan}
\affiliation{Spin Quantum Rectification Project, ERATO, Japan Science and Technology Agency, Sendai 980-8577, Japan}
\author{Zhiyong Qiu}
\email{qiuzy@imr.tohoku.ac.jp}\affiliation{WPI Advanced Institute for Materials Research, Tohoku University, Sendai 980-8577, Japan}
\affiliation{Spin Quantum Rectification Project, ERATO, Japan Science and Technology Agency, Sendai 980-8577, Japan}
\author{Joseph Barker}
\affiliation{Institute for Materials Research, Tohoku University, Sendai 980-8577, Japan}
\author{Koji Sato}
\affiliation{WPI Advanced Institute for Materials Research, Tohoku University, Sendai 980-8577, Japan}
\author{Kei Yamamoto}
\affiliation{Institute for Materials Research, Tohoku University, Sendai 980-8577, Japan}
\affiliation{Institut f\"ur Physik, Johannes Gutenberg Universit\"at Mainz, D-55099 Mainz, Germany}
\affiliation{Department of Physics, Kobe University, 1-1 Rokkodai, Kobe 657-8501, Japan}
\author{Sa\"ul V\'{e}lez}
\affiliation{CIC nanoGUNE, 20018 Donostia-San Sebastian, Basque Country, Spain}
\author{Juan M. Gomez-Perez}
\affiliation{CIC nanoGUNE, 20018 Donostia-San Sebastian, Basque Country, Spain}
\author{Luis E. Hueso}
\affiliation{CIC nanoGUNE, 20018 Donostia-San Sebastian, Basque Country, Spain}
\affiliation{IKERBASQUE, Basque Foundation for Science, 48011 Bilbao, Basque Country, Spain}
\author{F\`{e}lix Casanova}
\affiliation{CIC nanoGUNE, 20018 Donostia-San Sebastian, Basque Country, Spain}
\affiliation{IKERBASQUE, Basque Foundation for Science, 48011 Bilbao, Basque Country, Spain}
\author{Eiji Saitoh}
\affiliation{WPI Advanced Institute for Materials Research, Tohoku University, Sendai 980-8577, Japan}
\affiliation{Spin Quantum Rectification Project, ERATO, Japan Science and Technology Agency, Sendai 980-8577, Japan}
\affiliation{Institute for Materials Research, Tohoku University, Sendai 980-8577, Japan}
\affiliation{Advanced Science Research Center, Japan Atomic Energy Agency, Tokai 319-1195, Japan}

\begin{abstract}


Spin Hall magnetoresistance (SMR) has been investigated in Pt/NiO/YIG structures in a wide range of temperature and NiO thickness. The SMR shows a negative sign below a temperature which increases with the NiO thickness. This is contrary to a conventional SMR theory picture applied to Pt/YIG bilayer which always predicts a positive SMR. The negative SMR is found to presist even when NiO blocks the spin transmission between Pt and YIG, indicating it is governed by the spin current response of NiO layer. We explain the negative SMR by the NiO 'spin-flop' coupled with YIG, which can be overridden at higher temperature by positive SMR contribution from YIG. This highlights the role of magnetic structure in antiferromagnets for transport of pure spin current in multilayers.







\end{abstract}

\maketitle

Magnetoresistance plays essential roles in providing both a fundamental understanding of electron transport in magnetic materials and in various technological applications. Anisotropic magnetoresistance (AMR) \cite{campbell70,mcguire75}, giant magnetoresistance \cite{binaschPRB89,
baibichPRL88}, and tunneling magnetoresistance \cite{julliere75,miyazakiJMMM95,yuasaNatMater04,parkinNatMater04} underpin\ technologies in sensors, memories, and data storage. Recent studies of thin film bilayer systems comprised of a normal metal (NM) and a ferromagnetic insulator (FI) revealed a new type of magnetoresistance called spin Hall magnetoresistance (SMR) \cite{nakayamaPRL13,chenPRB13,althammerPRB13}, originating from the interplay between the spin accumulation at the NM/FI interface and the magnetization of the FI layer. When the NM layer has a significant spin-orbit interaction, e.g. Pt, an in-plane charge current $\bm j_{\rm{c}}$ induces a spin current via the spin Hall effect, which in turn generates a spin accumulation near the NM/FI interface. At the same time, this spin accumulation is affected by the orientation of the magnetization in the ferromagnet. The conductivity of the NM layer is thus subject to a magnetization dependent modification to the leading order in $\theta_{\rm SHE}^2$, where $\theta_{\rm SHE}$ is the spin Hall angle in the NM layer. 

Since the discovery of SMR, experimental studies were instigated in various systems \cite{Weiler2013,Meyer2014,Iguchi2014,Schreier2015,Lotze2014,Marmion2014,Isasa2014,Avci2015}. The amplitude of  SMR is defined as the difference of the resistivities with an applied field, $H$, parallel ($\rho_{\parallel}$) and perpendicular ($\rho_{\perp}$) to  $\bm j_c$: $\rho_{\rm{SMR}}$=$\rho_{\parallel}- \rho_{\perp}$. This is predicted to be always positive because when $H \parallel \bm j_c$, the FI can absorb more spin current, by which the backflow required to ensure the stationary state is reduced at the FI/NM interface, in turn causing less secondary forward charge current, and therefore gives : $\rho _{\parallel} > \rho _{\perp }$ \cite{nakayamaPRL13,chenPRB13}. Positive $\rho_{\rm{SMR}}$ is found in most experimental observations. 

Very recently, a negative SMR ($\rho_{\parallel} < \rho_{\perp}$) was reported when an antiferromagnetic (AFM) insulator, in this case NiO, is inserted between Pt and YIG \cite{Shang2016}. The negative SMR was also found to revert to the conventional positive\ sign at higher temperatures. Signal contamination from other magnetoresistances such as AMR was excluded by a systematic field angle dependence measurement. This result\ challenges the present understanding of SMR. Since the SMR does not change its sign in the Pt/YIG bilayer structure, the NiO layer must be the cause. However, it is not clear why NiO should give a negative SMR\ since antiferromagnets are thought only to affect the efficiency of the spin communication between Pt and YIG \cite{0295-5075-108-5-57005,Wang2014,moriyama2015,Qiu2016,Frangou2016,Linweiwei2016}.

In this letter, we report the temperature dependence of SMR in Pt/NiO/YIG structures with different thicknesses of NiO. The temperature at which the SMR becomes negative is found to depend on the  NiO thickness. The anomalous negative SMR at low temperatures is explained from a `spin-flop' configuration whereby the N\'eel order of the NiO is perpendicularly coupled to the magnetization of YIG \cite{Koon_spinflop}. As the spin conductivity of NiO increases with increasing temperature \cite{Qiu2016,Frangou2016,Linweiwei2016}, the moments of the YIG beneath have an increasing influence on the total SMR signal. The positive SMR contribution from YIG competes the negative SMR from NiO. At the sign change, the competition leads to a vanishing SMR. Above, in the high temperature regime, the positive SMR\ of the YIG dominates. We introduce a phenomenological model to describe the competition between the positive and negative SMR contributions, which reproduces the NiO thickness dependent SMR sign change behaviors in Pt/NiO/YIG.

\begin{figure}[t]
  \includegraphics[width=1\linewidth]{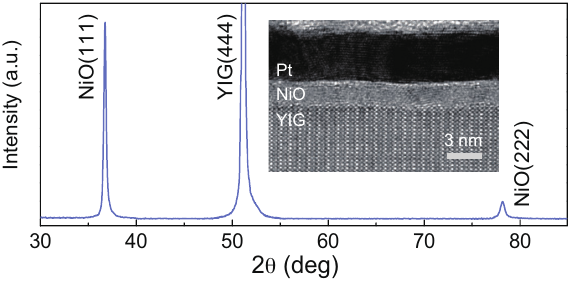}
\caption{ X-ray diffraction patterns of a 50 nm NiO film on YIG(111). Inset shows the cross section TEM photo for a Pt/NiO/YIG trilayer measured in the transport experiment.}
\label{fig0}
\end{figure}

An epitaxial YIG film of thickness 3 $\mu{\rm m}$ was grown on a gadolinium gallium garnet (111) substrate  prepared by the liquid phase epitaxy. NiO films of different thicknesses were grown by sputtering onto the YIG at 400 $^{\circ}$C. The film was then covered with 4 nm of sputtered Pt.  The X-ray diffraction patterns of a 50 nm NiO film on YIG is plotted in Fig.~\ref{fig0}, which only shows (111) and (222) NiO peaks of narrow line width. It suggests that the NiO film is of high crystallinity and a (111) preferred orientation. The inset in Figure~\ref{fig0} shows a representative cross-section TEM picture for a Pt/NiO/YIG sample, which confirms a good thickness uniformity and clean interface.

\begin{figure}[t]
\centering
  \includegraphics[width=1\linewidth]{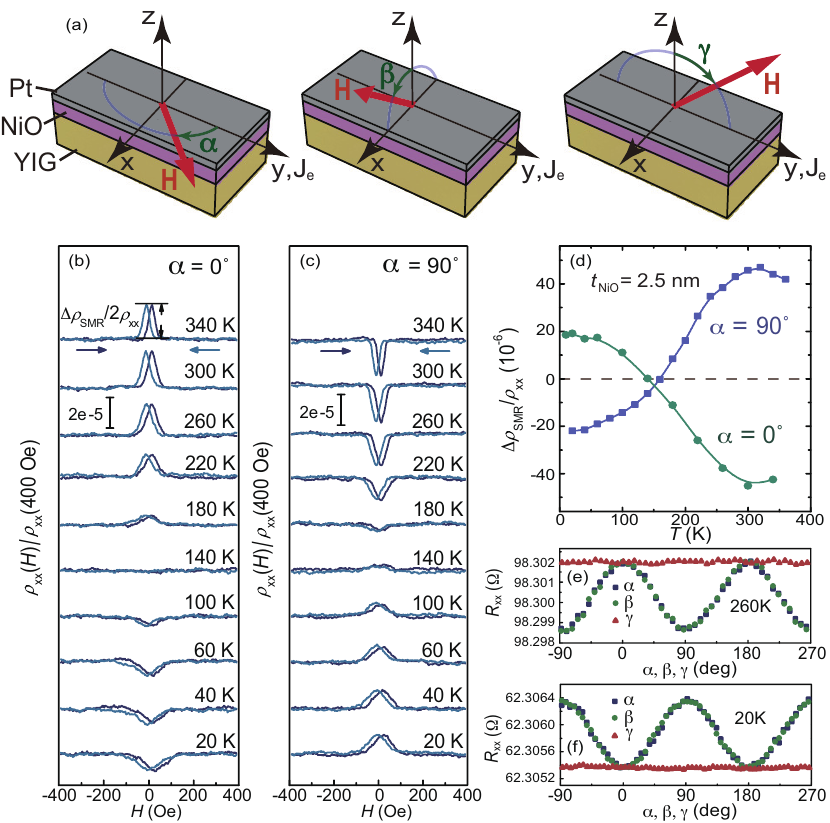}
  \caption{ (a),  The illustration for the magnetoresistance measurement setup for various magnetic field ($H$) orientations. $\alpha$, $\beta$ and $\gamma$ are the field angles defining the $H$ directions when $H$ is applied in the $x$-$y$, $x$-$z$ and $y$-$z$ planes, respectively. (b), (c), Magnetoresistance measured by $H$ sweeping for a Pt/NiO(2.5 nm)/YIG at $\alpha = 0^{\circ}$ and $90^{\circ}$ for various temperatures. (d), Temperature dependence of the SMR ratio $\Delta\rho_{\rm{SMR}}/\rho_{xx}$ for Pt/NiO(2.5 nm)/YIG at $\alpha = 0^{\circ}$ and $90^{\circ}$. (e), (f), Field angle dependent resistance measured for Pt/NiO(2 nm)/YIG at 260 K and 20 K with $|H|$= 20000 Oe, which shows positive and negative SMR, respectively.}\label{fig1}
\end{figure}

Figure~\ref{fig1}(a) shows the illustration of the magnetoresistance (MR) measurement setup and the definition of magnetic field angles. Standard four-probe method is employed for the MR observation at current density $\sim10^8$ A/$\rm m^2$, and MR can be detected either by sweeping $H$ along a fixed direction or by rotating $H$ of the same magnitude \cite{nakayamaPRL13}. Figure~\ref{fig1}(b) shows the MR measured by $H$ sweeping in a Pt/NiO(2.5 nm)/YIG sample at field angle $\alpha= 0^{\circ}$ for various temperatures. The range of magnetic field over which the magnetoresistance occurs, coincides with that of the switching process of YIG \cite{uchida_hight_T_SMR}. The MR data for $T >$140 K is consistent with the prediction $\rho_{\parallel} > \rho_{\perp}$ of the SMR theory. When $T$ =140 K, the MR nearly vanishes. For $T <$140 K, a sign change of MR is observed and the MR amplitude increases with decreasing temperature. The MR data from the same sample at field angle $\alpha= 90^{\circ}$ is plotted in Fig.~\ref{fig1}(c), which shows the same feature of the sign change. The SMR ratio $\Delta\rho_{\rm{SMR}}/\rho_{xx}$ extracted from Fig.~\ref{fig1}(b) and \ref{fig1}(c) are plotted in Fig.~\ref{fig1}(d). Figure~\ref{fig1}(e) and ~\ref{fig1}(f) show the field angle dependence of resistance in Pt/NiO(2 nm)/YIG at 260 K and 20 K, which not only reproduces the MR sign change behaviour, but confirms the SMR-type field angle dependence symmetry as well  \cite{Shang2016}. Thus, it looks reasonable to claim that SMR is the dominant contribution for the MR in Pt on NiO/YIG, since other mechanisms such as anisotropic magnetoresistance will cause a different field angle dependence \cite{ZhouMPEAMR2015}. However, the sign change of the magnetoresistance in the low temperature regime seems to be at odds with SMR which, conventionally, can only be positive \cite{chenPRB13}.

\begin{figure}[t]
\centering
  \includegraphics[width=1\linewidth]{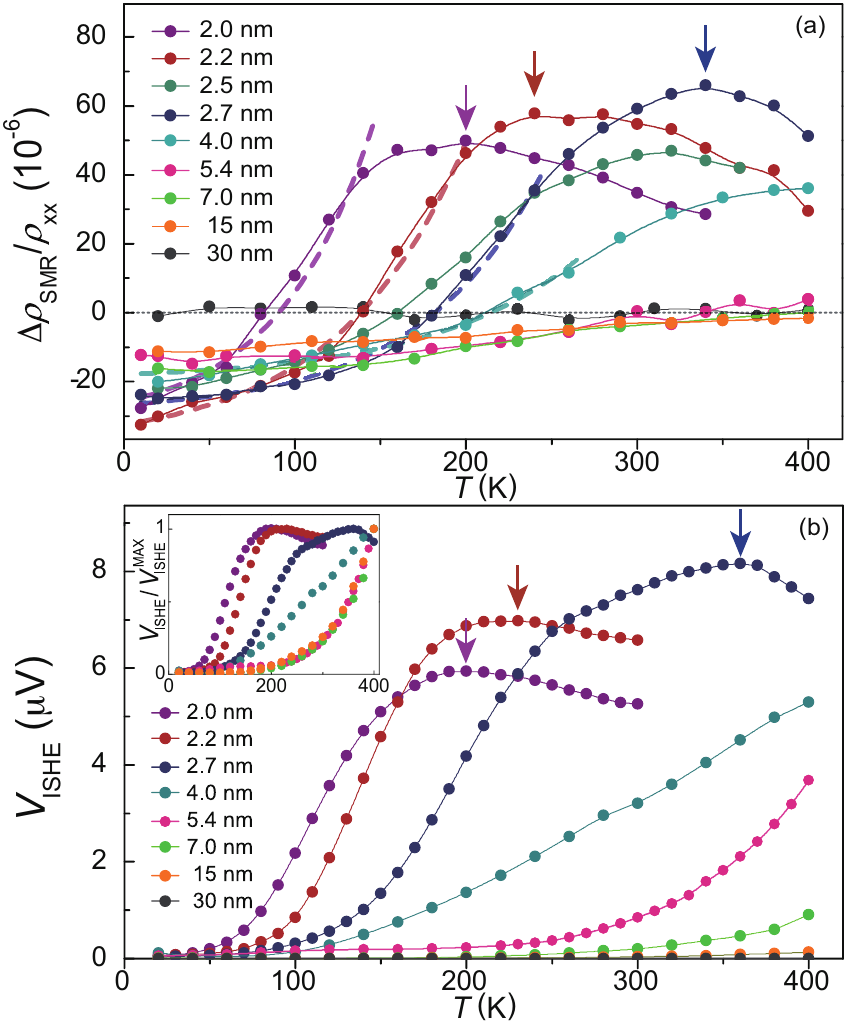}
  \caption{(a), The SMR ratio measured in Pt/NiO($d_{\rm{NiO}}$)/YIG devices with different NiO thickness $d_{\rm{NiO}}$ at various temperatures, which shows that the SMR sign change temperature is lower for a thinner NiO sample. The SMR ratio peak positions are marked by arrows. Negative SMR at low temperatures can be observed for all the NiO thickness except $d_{\rm{NiO}}$= 30 nm. The dashed curves are the fitting based on Eq. (2). (b), $V_{\rm{ISHE}}$ in Pt/NiO/YIG devices versus temperature from spin pumping measurement. The peak positions are marked by arrows, which are found to be close to the SMR ratio peak positions marked in Figure 2a. The inset shows the normalized $V_{\rm{ISHE}}$ temperature dependence. }\label{fig2}
\end{figure}

Fig.~\ref{fig2}(a) shows the temperature dependence of the SMR ratio measured in Pt/NiO/YIG devices with different NiO thicknesses, $d_{\rm{NiO}}$. The change in sign of the SMR occurs at higher temperatures in larger $d_{\rm{NiO}}$ samples. The $d_{\rm{NiO}}$ dependence proves to be a key piece of information for understanding the negative SMR. Furthermore, the SMR ratios have (positive) maxima at higher temperatures for thicker NiO samples. These $d_{\rm{NiO}}$ dependent characteristics show a quantitative effect of the NiO on the SMR\ modulation, rather than a nuanced interface effect \cite{Velez}.

To gain further insight into the temperature dependence of spin transport in NiO, we carried out spin pumping measurements for the same samples, in which spin current is injected from YIG through NiO to generate a voltage in Pt via the inverse spin Hall effect (ISHE) \cite{Wang2014}. The Pt/NiO/YIG device is placed on a coplanar waveguide which serves as a 5 GHz microwave source at 14 dbm, and the details of the experimental setup can be found elsewhere \cite{Qiu2016}. The ISHE voltage $V_{\rm{ISHE}}$ from all the samples is plotted against $T$ in Fig.~\ref{fig2}(b), the behaviour of which is very similar to the result we found in Pt/CoO/YIG \cite{Qiu2016}: spin transmission is nearly zero for low temperature limit and increases with temperature to reach the maximum around the N\'eel point. At room temperature, $V_{\rm{ISHE}}$ shows a non-monotonic $d_{\rm{NiO}}$ dependence, which is consistent with previous result. Fig.~\ref{fig2}(b) inset shows the normalized $V_{\rm{ISHE}}$ temperature dependence, in which the data for $d_{\rm{NiO}}=$ 5.4 nm, 7 nm and 15 nm collapse into a single curve. This confirms that the $V_{\rm{ISHE}}$ is governed by the NiO spin conductivity, which shows the same $T$ dependence when NiO is thick enough to exhibit bulk property. For $d_{\rm{NiO}}= 30$ nm, $V_{\rm{ISHE}}$ is below our measurement sensitivity 5 nV.

An important conclusion can be drawn by combining the results from SMR and spin pumping measurements: the negative SMR does not rely on the spin transmission between Pt and YIG, because it reaches the largest magnitude for the lowest temperature at which NiO spin conductivity vanishes. This argument can be further supported by the fact that the negative SMR is present even for $d_{\rm{NiO}}$= 15 nm, where the NiO spin conductivity is nearly zero throughout the entire temperature range. It indicates that the negative SMR is not caused by the magnetic moment of the YIG layer but that of the NiO layer, which is beyond any model based on spin communication between YIG and Pt \cite{chenPRB13,wwlin2017}.

\begin{figure}[t]
\centering
  \includegraphics[width=1\linewidth]{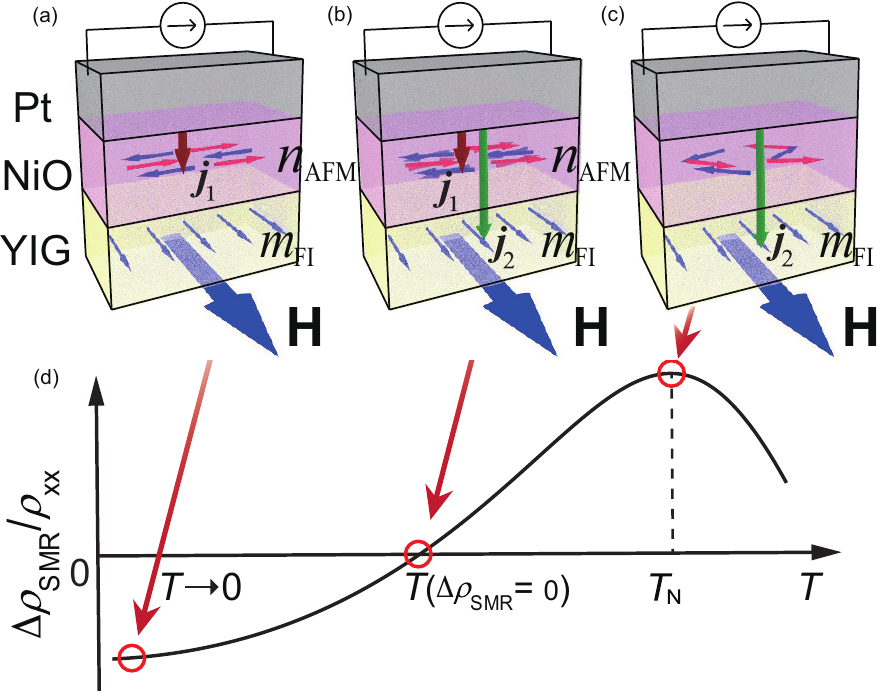}
  \caption{Illustrations for the magnetic structure and spin transport in Pt/NiO/YIG at different temperatures. The red and green arrows represent the phenomenologically described spin currents, $\bm j_1$ and $\bm j_2$ in Eq. (1), respectively. The length of the arrow describes the penetration depth of the spin current.  (a), $T$ close to the low temperature limit. (b), $T$ far above the low temperature limit and lower than the N\'eel temperature. (c),  $T$ higher than the N\'eel temperature. (d), Illustration of $T$-dependent SMR  in which the temperatures corresponding to the conditions in Fig. 4(a), (b) and (c) are marked with red circle.  }\label{fig3}
\end{figure}

Let us next provide an explanation for the negative SMR. The SMR in the trilayer system in this experiment is governed by the spin current through the Pt/NiO interface, which also reflects the effect of the presence of the NiO/YIG interface. The sign change and the thickness dependent behavior can be understood by assuming a `spin-flop' coupling between NiO and YIG \cite{Koon_spinflop,Schulthess1998}, which means the antiferromagnetic axis (N\'eel vector unit  $\bm n_{\rm{AFM}}$) in NiO is  perpendicular to the YIG magnetization unit vector $\bm m_{\rm{FI}}$ as illustrated in Fig.~\ref{fig3}(a). Although  a perpendicular coupling has not yet been confirmed experimentally for NiO on YIG, spin-flop coupling between NiO and other ferromagnets is quite common and well understood\cite{Koon_spinflop,Krug2008,LiJia2014}. For $d_{\rm{NiO}}$ below the domain wall width of NiO ($\sim$ 15 nm) \cite{NiOdomainsize}, which is the case for nearly all the samples, $\bm n_{\rm{AFM}}$ tends to be uniform in NiO, which is strongly coupled with YIG and can be manipulated by magnetic field \cite{NiOFeXMLD}. Thus, $\bm n_{\rm{AFM}}$ is always perpendicular to $H$ below the N\'eel temperature, because the $\bm m_{\rm{FI}}$ is parallel to $H$. In the low temperature limit, e.g. 10 K, the spin current generated in Pt can not penetrate through the NiO, thus the SMR signal is only caused by the NiO layer. The NiO local moments perpendicular to $H$ gives rise to a 90-degree phase shift in the SMR field angular dependence with respect to the conventional SMR \cite{nakayamaPRL13}. Such a 90-degree phase shift in a four-fold SMR field angular dependence is equivalent to a sign reversal in the conventional definition of MR, which explains the negative SMR in Pt/NiO/YIG at low temperatures. For $d_{\rm{NiO}}$= 30 nm which is beyond the domain wall width, $\bm n_{\rm{AFM}}$ at the Pt/NiO interface decouples with $\bm m_{\rm{FI}}$ and does not respond to $H$, which explains the vanishing of the negative SMR. 

At higher temperatures, but below the N\'eel point, antiferromagnetic order is maintained but the spin current from Pt has some transmission through NiO, which makes the effect of the YIG more visible as illustrated in Fig. 4(b). The negative SMR contribution from NiO and positive SMR contribution from YIG compete with each other. With increasing temperature, NiO becomes more transparent to the spin current, so the SMR contribution from YIG is enhanced. The SMR from NiO may also be suppressed because of the attenuation of the antiferromagnetic order at elevated temperatrues. As a result, the zero point of the SMR occurs at a temperature where the antiferromagnet is still in the ordered phase. Thinner NiO layers have a lower N\'eel point due to the finite size effect \cite{Alders1998}, hence the SMR also changes the sign at lower temperatures in thinner-NiO samples, which is in accordance with our observation shown in Fig. 3(a).

Around the N\'eel point as illustrated in Fig.~\ref{fig3}(c), the spin transparency of NiO are maximized \cite{Qiu2016}, where the SMR contribution from YIG reaches its peak value and the SMR contribution from NiO vanishes. As explained above, all the main features of the SMR data in Pt/NiO/YIG, such as negative SMR at low temperatures, $d_{\rm{NiO}}$ dependent sign change temperature and peak temperature, can be interpreted by the `spin-flop' configuration. Figure 4(d) shows an illustration of $\rho_{\rm SMR}$ temperature dependence, in which the temperature corresponding to these features are marked. We note that negative SMR has also been reported in bilayers of Pt on gadolinium iron garnet and Ar-sputtered YIG, in which the garnet interface moments can align perpendicularly to $H$ \cite{Velez,Granzhorn2016}.


A simple phenomenological model based on the picture discussed above can also provide a quantitative description of the observed SMR temperature dependence. 
 Let us consider a NM/AFM/FI trilayer system. 
The key assumption is that we can describe the spin current through the NM/AFM interface by
\begin{eqnarray}\label{spin_current}
e\bm j_s&=&G_{AF}\bm n_{\rm AFM}\times\left(\bm n_{\rm AFM}\times\bm\mu_s\right)+t(T)\bm m_{\rm FI}\times\left(\bm m_{\rm FI}\times\bm\mu_s\right) \nonumber\\
&=&e\bm j_1+e\bm j_2 ,
\end{eqnarray}
$G_{AF}$ is the real part of the spin mixing conductance at NM/AFM interface. $\bm\mu_s$ is the spin accumulation at the same interface. The first term, which we denote by $e\bm j_1$, is what is expected for NM/AFM bilayer systems as seen in the case studied in Ref.~\cite{takeiPRB14}. We have introduced the second term, which is denoted by $e\bm j_2$, to phenomenologically capture the effect of the FI layer. $t(T)$ encapsulates the temperature dependent transparency of the AFM to the spin current. In the case that the AFM is completely transparent the NM/FI bilayer result  $\bm m_{\rm FI}\times\left(\bm
m_{\rm FI}\times\bm\mu_s\right)$ is recovered. The linear combination of the NM/AFM and NM/FI terms has been chosen in an attempt to emulate our SMR data in the NM/AFM/FI system, seen in Fig.~\ref{fig2}(a), which seems to indicate a crossover from NM/AFM bilayer like behavior at low temperatures to NM/FI bilayer like behavior for higher temperatures.

Once we admit the form of the interfacial spin current in Eq.~(\ref{spin_current}), we can calculate the SMR by employing the diffusion equation and the Onsagar principle, according to Refs.~\cite{chenPRB13,*chenJPCM16}. The SMR contribution to the longitudinal resistivity then is given by
\begin{widetext}
\begin{equation}\label{eq:rho_smr}
\frac{\delta{\rho}}{\rho_0}=\frac{2\theta_{SHE}^2\lambda_N^2}{d_N\sigma}\frac{G_{AF}\cos^2\phi_n+t(T)\cos^2\phi_m+\nu t(T)G_{AF}\sin^2(\phi_m-\phi_n)}{1+\nu G_{AF}+\nu t(T)+ \nu ^2t(T)G_{AF}\sin^2(\phi_m-\phi_n)}\tanh^2\left(\frac{d_N}{2\lambda_N}\right)\,,
\end{equation}
\end{widetext}
where we defined $\nu =(2\lambda_N/\sigma)\coth(d_N/\lambda_N)$ with $\lambda_N$ and $\theta_{SHE}$ being the spin diffusion length and the spin Hall angle in NM, respectively, and $\sigma=\rho_0^{-1}$ is the conductivity of the NM layer. Here, $\phi_{n(m)}$ denotes the angle between $\bm n_{\rm AFM}(\bm m_{\rm FI})$ and the applied current $\bm j_c$ in NM. Now we set out a hypothesis that the crossover between the negative and positive SMR is of the same origin as the temperature dependence of the spin pumping signal (Fig.~\ref{fig2}(b)). In order to support it, the temperature dependence of $t(T)$ is obtained by fitting to the spin pumping data. The resulting function is then used alongside the other parameters in Eq.~(\ref{spin_current})  to fit the SMR data to test the validity of our model.

Based on the observation that the ISHE signal in Fig.~\ref{fig2}(b) is roughly exponential in the intermediate temperature regime, we employ $V_{\rm ISHE} \propto t(T) \propto e^{aT}-1$ to reproduce the temperature dependence of both spin pumping and SMR. The exponential behavior may not apply near the N\'eel temperature and the data points near and above the N\'eel temperature have been excluded from the fitting. Under these assumptions, $a$ can be determined from the spin pumping data (TABLE~\ref{table}).

We then fit $\delta \rho /\rho | _{ \phi _m = 0}  -\delta \rho /\rho |_{\phi _m = \pi /2} $ based on Eq.~(\ref{eq:rho_smr}) to the experimentally obtained SMR ratio $\Delta\rho_{\rm SMR}/\rho_{xx}$ in Fig.~\ref{fig2}(a) using the fitted value of $a$ from the $V_{\rm ISHE}$ data.
We fix $\lambda _N=1.5 {\rm nm}$, $ d_N= 4.0 {\rm nm}$, $\rho_0= \sigma^{-1}=$ 860 $\Omega{\rm nm}$, and $\theta _{SHE}= 0.05$, which are taken to be relevant values to the present experiment, and we further determine $G_{AF}$ and $G_F $ from the data, where the latter two are defined by $t(T)= G_F (e^{aT}-1) , \phi _n- \phi _m = \pi / 2$, respectively. The temperature dependence of $\rho _0$ and $\theta _{SHE}$ is ignored since they scale in some powers of $T$, which is wiped out by the exponential change in $t\left( T \right)$. The fitting curves can quantitatively reproduce the SMR sign change behavior as shown in Fig.~\ref{fig2}(a), and the fitting parameters are summarized in TABLE~\ref{table}. 
\begin{table}[h]
\begin{ruledtabular}
\begin{tabular}{c|c|c|c|c}
$d_{\rm{NiO}}$ & $a [K^{-1}] \times 10^{2}$ & $G_{AF}$ & $G_F$  \\
\hline 
$2.0$ & $1.83\pm0.22$ & $3.58\pm0.32 \times 10^{12}$  &$8.39\pm0.57 \times 10^{11}$  \\
$2.2$ & $1.38\pm0.19$ & $4.48\pm0.17 \times 10^{12}$  &$7.78\pm0.26 \times 10^{11}$  \\
$ 2.7 $ & $1.42\pm0.10$ &$ 3.67\pm0.09 \times 10^{12} $ & $3.01\pm0.08 \times 10^{11} $ \\
$4.0 $ & $1.16\pm0.09$ & $2.46\pm0.13 \times 10^{12} $ & $2.22\pm0.14 \times 10^{11} $  \\
\end{tabular}
\caption{The results of the fitting with the data from the SMR and spin pumping signals. The parameters are defined in the main text. The units of the last two columns are both $[ \Omega ^{-1} {\rm m}^{-2}]$. }\label{table}
\end{ruledtabular}
\end{table}


Our result highlights the importance of magnetic structure in AFM for spin transport, which suggests an alternative degree of freedom of spin manipulation. The NiO-induced SMR indicates that spin current response of AFM is anisotropic, which opens the possibility to use AFM insulator as a spin current valve or memory. 

$Note$ $added.$---Recently, we became aware of similar results for the SMR sign change observed in Pt/NiO/YIG by W. Lin $et$ $al.$ \cite{wwlin2017}. The NiO-thickness dependent SMR at room temperature was also reported by Yu-Ming Hung $et$ $al.$ \cite{SMR_RT_Kent}.

\bibliographystyle{apsrev4-1}

\begin{thebibliography}{41}%
\makeatletter
\providecommand \@ifxundefined [1]{%
 \@ifx{#1\undefined}
}%
\providecommand \@ifnum [1]{%
 \ifnum #1\expandafter \@firstoftwo
 \else \expandafter \@secondoftwo
 \fi
}%
\providecommand \@ifx [1]{%
 \ifx #1\expandafter \@firstoftwo
 \else \expandafter \@secondoftwo
 \fi
}%
\providecommand \natexlab [1]{#1}%
\providecommand \enquote  [1]{``#1''}%
\providecommand \bibnamefont  [1]{#1}%
\providecommand \bibfnamefont [1]{#1}%
\providecommand \citenamefont [1]{#1}%
\providecommand \href@noop [0]{\@secondoftwo}%
\providecommand \href [0]{\begingroup \@sanitize@url \@href}%
\providecommand \@href[1]{\@@startlink{#1}\@@href}%
\providecommand \@@href[1]{\endgroup#1\@@endlink}%
\providecommand \@sanitize@url [0]{\catcode `\\12\catcode `\$12\catcode
  `\&12\catcode `\#12\catcode `\^12\catcode `\_12\catcode `\%12\relax}%
\providecommand \@@startlink[1]{}%
\providecommand \@@endlink[0]{}%
\providecommand \url  [0]{\begingroup\@sanitize@url \@url }%
\providecommand \@url [1]{\endgroup\@href {#1}{\urlprefix }}%
\providecommand \urlprefix  [0]{URL }%
\providecommand \Eprint [0]{\href }%
\providecommand \doibase [0]{http://dx.doi.org/}%
\providecommand \selectlanguage [0]{\@gobble}%
\providecommand \bibinfo  [0]{\@secondoftwo}%
\providecommand \bibfield  [0]{\@secondoftwo}%
\providecommand \translation [1]{[#1]}%
\providecommand \BibitemOpen [0]{}%
\providecommand \bibitemStop [0]{}%
\providecommand \bibitemNoStop [0]{.\EOS\space}%
\providecommand \EOS [0]{\spacefactor3000\relax}%
\providecommand \BibitemShut  [1]{\csname bibitem#1\endcsname}%
\let\auto@bib@innerbib\@empty
\bibitem [{\citenamefont {Campbell}\ \emph {et~al.}(1970)\citenamefont
  {Campbell}, \citenamefont {Fert},\ and\ \citenamefont {Jaoul}}]{campbell70}%
  \BibitemOpen
  \bibfield  {author} {\bibinfo {author} {\bibfnamefont {I.~A.}\ \bibnamefont
  {Campbell}}, \bibinfo {author} {\bibfnamefont {A.}~\bibnamefont {Fert}}, \
  and\ \bibinfo {author} {\bibfnamefont {O.}~\bibnamefont {Jaoul}},\ }\href
  {http://stacks.iop.org/0022-3719/3/i=1S/a=310} {\bibfield  {journal}
  {\bibinfo  {journal} {J. Phys. C}\ }\textbf {\bibinfo {volume} {3}},\
  \bibinfo {pages} {S95} (\bibinfo {year} {1970})}\BibitemShut {NoStop}%
\bibitem [{\citenamefont {McGuire}\ and\ \citenamefont
  {Potter}(1975)}]{mcguire75}%
  \BibitemOpen
  \bibfield  {author} {\bibinfo {author} {\bibfnamefont {T.}~\bibnamefont
  {McGuire}}\ and\ \bibinfo {author} {\bibfnamefont {R.}~\bibnamefont
  {Potter}},\ }\href {\doibase 10.1109/TMAG.1975.1058782} {\bibfield  {journal}
  {\bibinfo  {journal} {IEEE Trans. Magn.}\ }\textbf {\bibinfo {volume} {11}},\
  \bibinfo {pages} {1018} (\bibinfo {year} {1975})}\BibitemShut {NoStop}%
\bibitem [{\citenamefont {Binasch}\ \emph {et~al.}(1989)\citenamefont
  {Binasch}, \citenamefont {Gr\"unberg}, \citenamefont {Saurenbach},\ and\
  \citenamefont {Zinn}}]{binaschPRB89}%
  \BibitemOpen
  \bibfield  {author} {\bibinfo {author} {\bibfnamefont {G.}~\bibnamefont
  {Binasch}}, \bibinfo {author} {\bibfnamefont {P.}~\bibnamefont {Gr\"unberg}},
  \bibinfo {author} {\bibfnamefont {F.}~\bibnamefont {Saurenbach}}, \ and\
  \bibinfo {author} {\bibfnamefont {W.}~\bibnamefont {Zinn}},\ }\href {\doibase
  10.1103/PhysRevB.39.4828} {\bibfield  {journal} {\bibinfo  {journal} {Phys.
  Rev. B}\ }\textbf {\bibinfo {volume} {39}},\ \bibinfo {pages} {4828}
  (\bibinfo {year} {1989})}\BibitemShut {NoStop}%
\bibitem [{\citenamefont {Baibich}\ \emph {et~al.}(1988)\citenamefont
  {Baibich}, \citenamefont {Broto}, \citenamefont {Fert}, \citenamefont
  {Van~Dau}, \citenamefont {Petroff}, \citenamefont {Etienne}, \citenamefont
  {Creuzet}, \citenamefont {Friederich},\ and\ \citenamefont
  {Chazelas}}]{baibichPRL88}%
  \BibitemOpen
  \bibfield  {author} {\bibinfo {author} {\bibfnamefont {M.~N.}\ \bibnamefont
  {Baibich}}, \bibinfo {author} {\bibfnamefont {J.~M.}\ \bibnamefont {Broto}},
  \bibinfo {author} {\bibfnamefont {A.}~\bibnamefont {Fert}}, \bibinfo {author}
  {\bibfnamefont {F.~N.}\ \bibnamefont {Van~Dau}}, \bibinfo {author}
  {\bibfnamefont {F.}~\bibnamefont {Petroff}}, \bibinfo {author} {\bibfnamefont
  {P.}~\bibnamefont {Etienne}}, \bibinfo {author} {\bibfnamefont
  {G.}~\bibnamefont {Creuzet}}, \bibinfo {author} {\bibfnamefont
  {A.}~\bibnamefont {Friederich}}, \ and\ \bibinfo {author} {\bibfnamefont
  {J.}~\bibnamefont {Chazelas}},\ }\href {\doibase 10.1103/PhysRevLett.61.2472}
  {\bibfield  {journal} {\bibinfo  {journal} {Phys. Rev. Lett.}\ }\textbf
  {\bibinfo {volume} {61}},\ \bibinfo {pages} {2472} (\bibinfo {year}
  {1988})}\BibitemShut {NoStop}%
\bibitem [{\citenamefont {Julliere}(1975)}]{julliere75}%
  \BibitemOpen
  \bibfield  {author} {\bibinfo {author} {\bibfnamefont {M.}~\bibnamefont
  {Julliere}},\ }\href {\doibase
  http://dx.doi.org/10.1016/0375-9601(75)90174-7} {\bibfield  {journal}
  {\bibinfo  {journal} {Phys. Lett. A}\ }\textbf {\bibinfo {volume} {54}},\
  \bibinfo {pages} {225 } (\bibinfo {year} {1975})}\BibitemShut {NoStop}%
\bibitem [{\citenamefont {Miyazaki}\ and\ \citenamefont
  {Tezuka}(1995)}]{miyazakiJMMM95}%
  \BibitemOpen
  \bibfield  {author} {\bibinfo {author} {\bibfnamefont {T.}~\bibnamefont
  {Miyazaki}}\ and\ \bibinfo {author} {\bibfnamefont {N.}~\bibnamefont
  {Tezuka}},\ }\href {\doibase http://dx.doi.org/10.1016/0304-8853(95)90001-2}
  {\bibfield  {journal} {\bibinfo  {journal} {Journal of Magnetism and Magnetic
  Materials}\ }\textbf {\bibinfo {volume} {139}},\ \bibinfo {pages} {L231 }
  (\bibinfo {year} {1995})}\BibitemShut {NoStop}%
\bibitem [{\citenamefont {Yuasa}\ \emph {et~al.}(2004)\citenamefont {Yuasa},
  \citenamefont {Nagahama}, \citenamefont {Fukushima}, \citenamefont {Suzuki},\
  and\ \citenamefont {Ando}}]{yuasaNatMater04}%
  \BibitemOpen
  \bibfield  {author} {\bibinfo {author} {\bibfnamefont {S.}~\bibnamefont
  {Yuasa}}, \bibinfo {author} {\bibfnamefont {T.}~\bibnamefont {Nagahama}},
  \bibinfo {author} {\bibfnamefont {A.}~\bibnamefont {Fukushima}}, \bibinfo
  {author} {\bibfnamefont {Y.}~\bibnamefont {Suzuki}}, \ and\ \bibinfo {author}
  {\bibfnamefont {K.}~\bibnamefont {Ando}},\ }\href {\doibase 10.1038/nmat1257}
  {\bibfield  {journal} {\bibinfo  {journal} {Nat. Mater.}\ }\textbf {\bibinfo
  {volume} {3}},\ \bibinfo {pages} {868} (\bibinfo {year} {2004})}\BibitemShut
  {NoStop}%
\bibitem [{\citenamefont {Parkin}\ \emph {et~al.}(2004)\citenamefont {Parkin},
  \citenamefont {Kaiser}, \citenamefont {Panchula}, \citenamefont {Rice},
  \citenamefont {Hughes}, \citenamefont {Samant},\ and\ \citenamefont
  {Yang}}]{parkinNatMater04}%
  \BibitemOpen
  \bibfield  {author} {\bibinfo {author} {\bibfnamefont {S.~S.~P.}\
  \bibnamefont {Parkin}}, \bibinfo {author} {\bibfnamefont {C.}~\bibnamefont
  {Kaiser}}, \bibinfo {author} {\bibfnamefont {A.}~\bibnamefont {Panchula}},
  \bibinfo {author} {\bibfnamefont {P.~M.}\ \bibnamefont {Rice}}, \bibinfo
  {author} {\bibfnamefont {B.}~\bibnamefont {Hughes}}, \bibinfo {author}
  {\bibfnamefont {M.}~\bibnamefont {Samant}}, \ and\ \bibinfo {author}
  {\bibfnamefont {S.-H.}\ \bibnamefont {Yang}},\ }\href {\doibase
  10.1038/nmat1256} {\bibfield  {journal} {\bibinfo  {journal} {Nat. Mater.}\
  }\textbf {\bibinfo {volume} {3}},\ \bibinfo {pages} {862} (\bibinfo {year}
  {2004})}\BibitemShut {NoStop}%
\bibitem [{\citenamefont {Nakayama}\ \emph {et~al.}(2013)\citenamefont
  {Nakayama}, \citenamefont {Althammer}, \citenamefont {Chen}, \citenamefont
  {Uchida}, \citenamefont {Kajiwara}, \citenamefont {Kikuchi}, \citenamefont
  {Ohtani}, \citenamefont {Gepr\"ags}, \citenamefont {Opel}, \citenamefont
  {Takahashi}, \citenamefont {Gross}, \citenamefont {Bauer}, \citenamefont
  {Goennenwein},\ and\ \citenamefont {Saitoh}}]{nakayamaPRL13}%
  \BibitemOpen
  \bibfield  {author} {\bibinfo {author} {\bibfnamefont {H.}~\bibnamefont
  {Nakayama}}, \bibinfo {author} {\bibfnamefont {M.}~\bibnamefont {Althammer}},
  \bibinfo {author} {\bibfnamefont {Y.-T.}\ \bibnamefont {Chen}}, \bibinfo
  {author} {\bibfnamefont {K.}~\bibnamefont {Uchida}}, \bibinfo {author}
  {\bibfnamefont {Y.}~\bibnamefont {Kajiwara}}, \bibinfo {author}
  {\bibfnamefont {D.}~\bibnamefont {Kikuchi}}, \bibinfo {author} {\bibfnamefont
  {T.}~\bibnamefont {Ohtani}}, \bibinfo {author} {\bibfnamefont
  {S.}~\bibnamefont {Gepr\"ags}}, \bibinfo {author} {\bibfnamefont
  {M.}~\bibnamefont {Opel}}, \bibinfo {author} {\bibfnamefont {S.}~\bibnamefont
  {Takahashi}}, \bibinfo {author} {\bibfnamefont {R.}~\bibnamefont {Gross}},
  \bibinfo {author} {\bibfnamefont {G.~E.~W.}\ \bibnamefont {Bauer}}, \bibinfo
  {author} {\bibfnamefont {S.~T.~B.}\ \bibnamefont {Goennenwein}}, \ and\
  \bibinfo {author} {\bibfnamefont {E.}~\bibnamefont {Saitoh}},\ }\href
  {\doibase 10.1103/PhysRevLett.110.206601} {\bibfield  {journal} {\bibinfo
  {journal} {Phys. Rev. Lett.}\ }\textbf {\bibinfo {volume} {110}},\ \bibinfo
  {pages} {206601} (\bibinfo {year} {2013})}\BibitemShut {NoStop}%
\bibitem [{\citenamefont {Chen}\ \emph {et~al.}(2013)\citenamefont {Chen},
  \citenamefont {Takahashi}, \citenamefont {Nakayama}, \citenamefont
  {Althammer}, \citenamefont {Goennenwein}, \citenamefont {Saitoh},\ and\
  \citenamefont {Bauer}}]{chenPRB13}%
  \BibitemOpen
  \bibfield  {author} {\bibinfo {author} {\bibfnamefont {Y.-T.}\ \bibnamefont
  {Chen}}, \bibinfo {author} {\bibfnamefont {S.}~\bibnamefont {Takahashi}},
  \bibinfo {author} {\bibfnamefont {H.}~\bibnamefont {Nakayama}}, \bibinfo
  {author} {\bibfnamefont {M.}~\bibnamefont {Althammer}}, \bibinfo {author}
  {\bibfnamefont {S.~T.~B.}\ \bibnamefont {Goennenwein}}, \bibinfo {author}
  {\bibfnamefont {E.}~\bibnamefont {Saitoh}}, \ and\ \bibinfo {author}
  {\bibfnamefont {G.~E.~W.}\ \bibnamefont {Bauer}},\ }\href {\doibase
  10.1103/PhysRevB.87.144411} {\bibfield  {journal} {\bibinfo  {journal} {Phys.
  Rev. B}\ }\textbf {\bibinfo {volume} {87}},\ \bibinfo {pages} {144411}
  (\bibinfo {year} {2013})}\BibitemShut {NoStop}%
\bibitem [{\citenamefont {Althammer}\ \emph {et~al.}(2013)\citenamefont
  {Althammer}, \citenamefont {Meyer}, \citenamefont {Nakayama}, \citenamefont
  {Schreier}, \citenamefont {Altmannshofer}, \citenamefont {Weiler},
  \citenamefont {Huebl}, \citenamefont {Gepr\"ags}, \citenamefont {Opel},
  \citenamefont {Gross}, \citenamefont {Meier}, \citenamefont {Klewe},
  \citenamefont {Kuschel}, \citenamefont {Schmalhorst}, \citenamefont {Reiss},
  \citenamefont {Shen}, \citenamefont {Gupta}, \citenamefont {Chen},
  \citenamefont {Bauer}, \citenamefont {Saitoh},\ and\ \citenamefont
  {Goennenwein}}]{althammerPRB13}%
  \BibitemOpen
  \bibfield  {author} {\bibinfo {author} {\bibfnamefont {M.}~\bibnamefont
  {Althammer}}, \bibinfo {author} {\bibfnamefont {S.}~\bibnamefont {Meyer}},
  \bibinfo {author} {\bibfnamefont {H.}~\bibnamefont {Nakayama}}, \bibinfo
  {author} {\bibfnamefont {M.}~\bibnamefont {Schreier}}, \bibinfo {author}
  {\bibfnamefont {S.}~\bibnamefont {Altmannshofer}}, \bibinfo {author}
  {\bibfnamefont {M.}~\bibnamefont {Weiler}}, \bibinfo {author} {\bibfnamefont
  {H.}~\bibnamefont {Huebl}}, \bibinfo {author} {\bibfnamefont
  {S.}~\bibnamefont {Gepr\"ags}}, \bibinfo {author} {\bibfnamefont
  {M.}~\bibnamefont {Opel}}, \bibinfo {author} {\bibfnamefont {R.}~\bibnamefont
  {Gross}}, \bibinfo {author} {\bibfnamefont {D.}~\bibnamefont {Meier}},
  \bibinfo {author} {\bibfnamefont {C.}~\bibnamefont {Klewe}}, \bibinfo
  {author} {\bibfnamefont {T.}~\bibnamefont {Kuschel}}, \bibinfo {author}
  {\bibfnamefont {J.-M.}\ \bibnamefont {Schmalhorst}}, \bibinfo {author}
  {\bibfnamefont {G.}~\bibnamefont {Reiss}}, \bibinfo {author} {\bibfnamefont
  {L.}~\bibnamefont {Shen}}, \bibinfo {author} {\bibfnamefont {A.}~\bibnamefont
  {Gupta}}, \bibinfo {author} {\bibfnamefont {Y.-T.}\ \bibnamefont {Chen}},
  \bibinfo {author} {\bibfnamefont {G.~E.~W.}\ \bibnamefont {Bauer}}, \bibinfo
  {author} {\bibfnamefont {E.}~\bibnamefont {Saitoh}}, \ and\ \bibinfo {author}
  {\bibfnamefont {S.~T.~B.}\ \bibnamefont {Goennenwein}},\ }\href {\doibase
  10.1103/PhysRevB.87.224401} {\bibfield  {journal} {\bibinfo  {journal} {Phys.
  Rev. B}\ }\textbf {\bibinfo {volume} {87}},\ \bibinfo {pages} {224401}
  (\bibinfo {year} {2013})}\BibitemShut {NoStop}%
\bibitem [{\citenamefont {Weiler}\ \emph {et~al.}(2013)\citenamefont {Weiler},
  \citenamefont {Althammer}, \citenamefont {Schreier}, \citenamefont {Lotze},
  \citenamefont {Pernpeintner}, \citenamefont {Meyer}, \citenamefont {Huebl},
  \citenamefont {Gross}, \citenamefont {Kamra}, \citenamefont {Xiao},
  \citenamefont {Chen}, \citenamefont {Jiao}, \citenamefont {Bauer},\ and\
  \citenamefont {Goennenwein}}]{Weiler2013}%
  \BibitemOpen
  \bibfield  {author} {\bibinfo {author} {\bibfnamefont {M.}~\bibnamefont
  {Weiler}}, \bibinfo {author} {\bibfnamefont {M.}~\bibnamefont {Althammer}},
  \bibinfo {author} {\bibfnamefont {M.}~\bibnamefont {Schreier}}, \bibinfo
  {author} {\bibfnamefont {J.}~\bibnamefont {Lotze}}, \bibinfo {author}
  {\bibfnamefont {M.}~\bibnamefont {Pernpeintner}}, \bibinfo {author}
  {\bibfnamefont {S.}~\bibnamefont {Meyer}}, \bibinfo {author} {\bibfnamefont
  {H.}~\bibnamefont {Huebl}}, \bibinfo {author} {\bibfnamefont
  {R.}~\bibnamefont {Gross}}, \bibinfo {author} {\bibfnamefont
  {A.}~\bibnamefont {Kamra}}, \bibinfo {author} {\bibfnamefont
  {J.}~\bibnamefont {Xiao}}, \bibinfo {author} {\bibfnamefont {Y.-T.}\
  \bibnamefont {Chen}}, \bibinfo {author} {\bibfnamefont {H.}~\bibnamefont
  {Jiao}}, \bibinfo {author} {\bibfnamefont {G.~E.~W.}\ \bibnamefont {Bauer}},
  \ and\ \bibinfo {author} {\bibfnamefont {S.~T.~B.}\ \bibnamefont
  {Goennenwein}},\ }\href {\doibase 10.1103/PhysRevLett.111.176601} {\bibfield
  {journal} {\bibinfo  {journal} {Phys. Rev. Lett.}\ }\textbf {\bibinfo
  {volume} {111}},\ \bibinfo {pages} {176601} (\bibinfo {year}
  {2013})}\BibitemShut {NoStop}%
\bibitem [{\citenamefont {Meyer}\ \emph {et~al.}(2014)\citenamefont {Meyer},
  \citenamefont {Althammer}, \citenamefont {Gepr\"ags}, \citenamefont {Opel},
  \citenamefont {Gross},\ and\ \citenamefont {Goennenwein}}]{Meyer2014}%
  \BibitemOpen
  \bibfield  {author} {\bibinfo {author} {\bibfnamefont {S.}~\bibnamefont
  {Meyer}}, \bibinfo {author} {\bibfnamefont {M.}~\bibnamefont {Althammer}},
  \bibinfo {author} {\bibfnamefont {S.}~\bibnamefont {Gepr\"ags}}, \bibinfo
  {author} {\bibfnamefont {M.}~\bibnamefont {Opel}}, \bibinfo {author}
  {\bibfnamefont {R.}~\bibnamefont {Gross}}, \ and\ \bibinfo {author}
  {\bibfnamefont {S.~T.~B.}\ \bibnamefont {Goennenwein}},\ }\href
  {http://scitation.aip.org/content/aip/journal/apl/104/24/10.1063/1.4885086}
  {\bibfield  {journal} {\bibinfo  {journal} {Applied Physics Letters}\
  }\textbf {\bibinfo {volume} {104}},\ \bibinfo {eid} {242411} (\bibinfo {year}
  {2014})}\BibitemShut {NoStop}%
\bibitem [{\citenamefont {Iguchi}\ \emph {et~al.}(2014)\citenamefont {Iguchi},
  \citenamefont {Sato}, \citenamefont {Hirobe}, \citenamefont {Daimon},\ and\
  \citenamefont {Saitoh}}]{Iguchi2014}%
  \BibitemOpen
  \bibfield  {author} {\bibinfo {author} {\bibfnamefont {R.}~\bibnamefont
  {Iguchi}}, \bibinfo {author} {\bibfnamefont {K.}~\bibnamefont {Sato}},
  \bibinfo {author} {\bibfnamefont {D.}~\bibnamefont {Hirobe}}, \bibinfo
  {author} {\bibfnamefont {S.}~\bibnamefont {Daimon}}, \ and\ \bibinfo {author}
  {\bibfnamefont {E.}~\bibnamefont {Saitoh}},\ }\href
  {http://stacks.iop.org/1882-0786/7/i=1/a=013003} {\bibfield  {journal}
  {\bibinfo  {journal} {Applied Physics Express}\ }\textbf {\bibinfo {volume}
  {7}},\ \bibinfo {pages} {013003} (\bibinfo {year} {2014})}\BibitemShut
  {NoStop}%
\bibitem [{\citenamefont {Schreier}\ \emph {et~al.}(2015)\citenamefont
  {Schreier}, \citenamefont {Chiba}, \citenamefont {Niedermayr}, \citenamefont
  {Lotze}, \citenamefont {Huebl}, \citenamefont {Gepr\"ags}, \citenamefont
  {Takahashi}, \citenamefont {Bauer}, \citenamefont {Gross},\ and\
  \citenamefont {Goennenwein}}]{Schreier2015}%
  \BibitemOpen
  \bibfield  {author} {\bibinfo {author} {\bibfnamefont {M.}~\bibnamefont
  {Schreier}}, \bibinfo {author} {\bibfnamefont {T.}~\bibnamefont {Chiba}},
  \bibinfo {author} {\bibfnamefont {A.}~\bibnamefont {Niedermayr}}, \bibinfo
  {author} {\bibfnamefont {J.}~\bibnamefont {Lotze}}, \bibinfo {author}
  {\bibfnamefont {H.}~\bibnamefont {Huebl}}, \bibinfo {author} {\bibfnamefont
  {S.}~\bibnamefont {Gepr\"ags}}, \bibinfo {author} {\bibfnamefont
  {S.}~\bibnamefont {Takahashi}}, \bibinfo {author} {\bibfnamefont {G.~E.~W.}\
  \bibnamefont {Bauer}}, \bibinfo {author} {\bibfnamefont {R.}~\bibnamefont
  {Gross}}, \ and\ \bibinfo {author} {\bibfnamefont {S.~T.~B.}\ \bibnamefont
  {Goennenwein}},\ }\href {\doibase 10.1103/PhysRevB.92.144411} {\bibfield
  {journal} {\bibinfo  {journal} {Phys. Rev. B}\ }\textbf {\bibinfo {volume}
  {92}},\ \bibinfo {pages} {144411} (\bibinfo {year} {2015})}\BibitemShut
  {NoStop}%
\bibitem [{\citenamefont {Lotze}\ \emph {et~al.}(2014)\citenamefont {Lotze},
  \citenamefont {Huebl}, \citenamefont {Gross},\ and\ \citenamefont
  {Goennenwein}}]{Lotze2014}%
  \BibitemOpen
  \bibfield  {author} {\bibinfo {author} {\bibfnamefont {J.}~\bibnamefont
  {Lotze}}, \bibinfo {author} {\bibfnamefont {H.}~\bibnamefont {Huebl}},
  \bibinfo {author} {\bibfnamefont {R.}~\bibnamefont {Gross}}, \ and\ \bibinfo
  {author} {\bibfnamefont {S.~T.~B.}\ \bibnamefont {Goennenwein}},\ }\href
  {\doibase 10.1103/PhysRevB.90.174419} {\bibfield  {journal} {\bibinfo
  {journal} {Phys. Rev. B}\ }\textbf {\bibinfo {volume} {90}},\ \bibinfo
  {pages} {174419} (\bibinfo {year} {2014})}\BibitemShut {NoStop}%
\bibitem [{\citenamefont {Marmion}\ \emph {et~al.}(2014)\citenamefont
  {Marmion}, \citenamefont {Ali}, \citenamefont {McLaren}, \citenamefont
  {Williams},\ and\ \citenamefont {Hickey}}]{Marmion2014}%
  \BibitemOpen
  \bibfield  {author} {\bibinfo {author} {\bibfnamefont {S.~R.}\ \bibnamefont
  {Marmion}}, \bibinfo {author} {\bibfnamefont {M.}~\bibnamefont {Ali}},
  \bibinfo {author} {\bibfnamefont {M.}~\bibnamefont {McLaren}}, \bibinfo
  {author} {\bibfnamefont {D.~A.}\ \bibnamefont {Williams}}, \ and\ \bibinfo
  {author} {\bibfnamefont {B.~J.}\ \bibnamefont {Hickey}},\ }\href {\doibase
  10.1103/PhysRevB.89.220404} {\bibfield  {journal} {\bibinfo  {journal} {Phys.
  Rev. B}\ }\textbf {\bibinfo {volume} {89}},\ \bibinfo {pages} {220404}
  (\bibinfo {year} {2014})}\BibitemShut {NoStop}%
\bibitem [{\citenamefont {Isasa}\ \emph {et~al.}(2014)\citenamefont {Isasa},
  \citenamefont {Bedoya-Pinto}, \citenamefont {V{\'{e}}lez}, \citenamefont
  {Golmar}, \citenamefont {SÃ¡nchez}, \citenamefont {Hueso}, \citenamefont
  {Fontcuberta},\ and\ \citenamefont {Casanova}}]{Isasa2014}%
  \BibitemOpen
  \bibfield  {author} {\bibinfo {author} {\bibfnamefont {M.}~\bibnamefont
  {Isasa}}, \bibinfo {author} {\bibfnamefont {A.}~\bibnamefont {Bedoya-Pinto}},
  \bibinfo {author} {\bibfnamefont {S.}~\bibnamefont {V{\'{e}}lez}}, \bibinfo
  {author} {\bibfnamefont {F.}~\bibnamefont {Golmar}}, \bibinfo {author}
  {\bibfnamefont {F.}~\bibnamefont {SÃ¡nchez}}, \bibinfo {author}
  {\bibfnamefont {L.~E.}\ \bibnamefont {Hueso}}, \bibinfo {author}
  {\bibfnamefont {J.}~\bibnamefont {Fontcuberta}}, \ and\ \bibinfo {author}
  {\bibfnamefont {F.}~\bibnamefont {Casanova}},\ }\href@noop {} {\bibfield
  {journal} {\bibinfo  {journal} {Applied Physics Letters}\ }\textbf {\bibinfo
  {volume} {105}},\ \bibinfo {eid} {142402} (\bibinfo {year}
  {2014})}\BibitemShut {NoStop}%
\bibitem [{\citenamefont {Avci}\ \emph {et~al.}(2015)\citenamefont {Avci},
  \citenamefont {Garello}, \citenamefont {Ghosh}, \citenamefont {Gabureac},
  \citenamefont {Alvarado},\ and\ \citenamefont {Gambardella}}]{Avci2015}%
  \BibitemOpen
  \bibfield  {author} {\bibinfo {author} {\bibfnamefont {C.~O.}\ \bibnamefont
  {Avci}}, \bibinfo {author} {\bibfnamefont {K.}~\bibnamefont {Garello}},
  \bibinfo {author} {\bibfnamefont {A.}~\bibnamefont {Ghosh}}, \bibinfo
  {author} {\bibfnamefont {M.}~\bibnamefont {Gabureac}}, \bibinfo {author}
  {\bibfnamefont {S.~F.}\ \bibnamefont {Alvarado}}, \ and\ \bibinfo {author}
  {\bibfnamefont {P.}~\bibnamefont {Gambardella}},\ }\href {\doibase
  10.1038/nphys3356} {\bibfield  {journal} {\bibinfo  {journal} {Nat Phys}\
  }\textbf {\bibinfo {volume} {11}},\ \bibinfo {pages} {570} (\bibinfo {year}
  {2015})}\BibitemShut {NoStop}%
\bibitem [{\citenamefont {Shang}\ \emph {et~al.}(2016)\citenamefont {Shang},
  \citenamefont {Zhan}, \citenamefont {Yang}, \citenamefont {Zuo},
  \citenamefont {Xie}, \citenamefont {Liu}, \citenamefont {Zhang},
  \citenamefont {Zhang}, \citenamefont {Li}, \citenamefont {Wang},
  \citenamefont {Wu}, \citenamefont {Zhang},\ and\ \citenamefont
  {Li}}]{Shang2016}%
  \BibitemOpen
  \bibfield  {author} {\bibinfo {author} {\bibfnamefont {T.}~\bibnamefont
  {Shang}}, \bibinfo {author} {\bibfnamefont {Q.~F.}\ \bibnamefont {Zhan}},
  \bibinfo {author} {\bibfnamefont {H.~L.}\ \bibnamefont {Yang}}, \bibinfo
  {author} {\bibfnamefont {Z.~H.}\ \bibnamefont {Zuo}}, \bibinfo {author}
  {\bibfnamefont {Y.~L.}\ \bibnamefont {Xie}}, \bibinfo {author} {\bibfnamefont
  {L.~P.}\ \bibnamefont {Liu}}, \bibinfo {author} {\bibfnamefont {S.~L.}\
  \bibnamefont {Zhang}}, \bibinfo {author} {\bibfnamefont {Y.}~\bibnamefont
  {Zhang}}, \bibinfo {author} {\bibfnamefont {H.~H.}\ \bibnamefont {Li}},
  \bibinfo {author} {\bibfnamefont {B.~M.}\ \bibnamefont {Wang}}, \bibinfo
  {author} {\bibfnamefont {Y.~H.}\ \bibnamefont {Wu}}, \bibinfo {author}
  {\bibfnamefont {S.}~\bibnamefont {Zhang}}, \ and\ \bibinfo {author}
  {\bibfnamefont {R.-W.}\ \bibnamefont {Li}},\ }\href
  {http://scitation.aip.org/content/aip/journal/apl/109/3/10.1063/1.4959573}
  {\bibfield  {journal} {\bibinfo  {journal} {Applied Physics Letters}\
  }\textbf {\bibinfo {volume} {109}},\ \bibinfo {eid} {032410} (\bibinfo {year}
  {2016})}\BibitemShut {NoStop}%
\bibitem [{\citenamefont {Hahn}\ \emph {et~al.}(2014)\citenamefont {Hahn},
  \citenamefont {de~Loubens}, \citenamefont {Naletov}, \citenamefont {Youssef},
  \citenamefont {Klein},\ and\ \citenamefont {Viret}}]{0295-5075-108-5-57005}%
  \BibitemOpen
  \bibfield  {author} {\bibinfo {author} {\bibfnamefont {C.}~\bibnamefont
  {Hahn}}, \bibinfo {author} {\bibfnamefont {G.}~\bibnamefont {de~Loubens}},
  \bibinfo {author} {\bibfnamefont {V.~V.}\ \bibnamefont {Naletov}}, \bibinfo
  {author} {\bibfnamefont {J.~B.}\ \bibnamefont {Youssef}}, \bibinfo {author}
  {\bibfnamefont {O.}~\bibnamefont {Klein}}, \ and\ \bibinfo {author}
  {\bibfnamefont {M.}~\bibnamefont {Viret}},\ }\href
  {http://stacks.iop.org/0295-5075/108/i=5/a=57005} {\bibfield  {journal}
  {\bibinfo  {journal} {EPL (Europhysics Letters)}\ }\textbf {\bibinfo {volume}
  {108}},\ \bibinfo {pages} {57005} (\bibinfo {year} {2014})}\BibitemShut
  {NoStop}%
\bibitem [{\citenamefont {Wang}\ \emph {et~al.}(2014)\citenamefont {Wang},
  \citenamefont {Du}, \citenamefont {Hammel},\ and\ \citenamefont
  {Yang}}]{Wang2014}%
  \BibitemOpen
  \bibfield  {author} {\bibinfo {author} {\bibfnamefont {H.}~\bibnamefont
  {Wang}}, \bibinfo {author} {\bibfnamefont {C.}~\bibnamefont {Du}}, \bibinfo
  {author} {\bibfnamefont {P.~C.}\ \bibnamefont {Hammel}}, \ and\ \bibinfo
  {author} {\bibfnamefont {F.}~\bibnamefont {Yang}},\ }\href {\doibase
  10.1103/PhysRevLett.113.097202} {\bibfield  {journal} {\bibinfo  {journal}
  {Phys. Rev. Lett.}\ }\textbf {\bibinfo {volume} {113}},\ \bibinfo {pages}
  {097202} (\bibinfo {year} {2014})}\BibitemShut {NoStop}%
\bibitem [{\citenamefont {Moriyama}\ \emph {et~al.}(2015)\citenamefont
  {Moriyama}, \citenamefont {Takei}, \citenamefont {Nagata}, \citenamefont
  {Yoshimura}, \citenamefont {Matsuzaki}, \citenamefont {Terashima},
  \citenamefont {Tserkovnyak},\ and\ \citenamefont {Ono}}]{moriyama2015}%
  \BibitemOpen
  \bibfield  {author} {\bibinfo {author} {\bibfnamefont {T.}~\bibnamefont
  {Moriyama}}, \bibinfo {author} {\bibfnamefont {S.}~\bibnamefont {Takei}},
  \bibinfo {author} {\bibfnamefont {M.}~\bibnamefont {Nagata}}, \bibinfo
  {author} {\bibfnamefont {Y.}~\bibnamefont {Yoshimura}}, \bibinfo {author}
  {\bibfnamefont {N.}~\bibnamefont {Matsuzaki}}, \bibinfo {author}
  {\bibfnamefont {T.}~\bibnamefont {Terashima}}, \bibinfo {author}
  {\bibfnamefont {Y.}~\bibnamefont {Tserkovnyak}}, \ and\ \bibinfo {author}
  {\bibfnamefont {T.}~\bibnamefont {Ono}},\ }\href
  {http://scitation.aip.org/content/aip/journal/apl/106/16/10.1063/1.4918990;jsessionid=cdjrfTs6Jsdh4WtxlHTdH03f.x-aip-live-03}
  {\bibfield  {journal} {\bibinfo  {journal} {Applied Physics Letters}\
  }\textbf {\bibinfo {volume} {106}},\ \bibinfo {eid} {162406} (\bibinfo {year}
  {2015})}\BibitemShut {NoStop}%
\bibitem [{\citenamefont {Qiu}\ \emph {et~al.}(2016)\citenamefont {Qiu},
  \citenamefont {Li}, \citenamefont {Hou}, \citenamefont {Arenholz},
  \citenamefont {N'Diaye}, \citenamefont {Tan}, \citenamefont {Uchida},
  \citenamefont {Sato}, \citenamefont {Okamoto}, \citenamefont {Tserkovnyak},
  \citenamefont {Qiu},\ and\ \citenamefont {Saitoh}}]{Qiu2016}%
  \BibitemOpen
  \bibfield  {author} {\bibinfo {author} {\bibfnamefont {Z.}~\bibnamefont
  {Qiu}}, \bibinfo {author} {\bibfnamefont {J.}~\bibnamefont {Li}}, \bibinfo
  {author} {\bibfnamefont {D.}~\bibnamefont {Hou}}, \bibinfo {author}
  {\bibfnamefont {E.}~\bibnamefont {Arenholz}}, \bibinfo {author}
  {\bibfnamefont {A.~T.}\ \bibnamefont {N'Diaye}}, \bibinfo {author}
  {\bibfnamefont {A.}~\bibnamefont {Tan}}, \bibinfo {author} {\bibfnamefont
  {K.-i.}\ \bibnamefont {Uchida}}, \bibinfo {author} {\bibfnamefont
  {K.}~\bibnamefont {Sato}}, \bibinfo {author} {\bibfnamefont {S.}~\bibnamefont
  {Okamoto}}, \bibinfo {author} {\bibfnamefont {Y.}~\bibnamefont
  {Tserkovnyak}}, \bibinfo {author} {\bibfnamefont {Z.~Q.}\ \bibnamefont
  {Qiu}}, \ and\ \bibinfo {author} {\bibfnamefont {E.}~\bibnamefont {Saitoh}},\
  }\href {http://dx.doi.org/10.1038/ncomms12670 http://10.1038/ncomms12670
  http://www.nature.com/articles/ncomms12670{\#}supplementary-information}
  {\bibfield  {journal} {\bibinfo  {journal} {Nature Communications}\ }\textbf
  {\bibinfo {volume} {7}},\ \bibinfo {pages} {12670} (\bibinfo {year}
  {2016})}\BibitemShut {NoStop}%
\bibitem [{\citenamefont {Frangou}\ \emph {et~al.}(2016)\citenamefont
  {Frangou}, \citenamefont {Oyarz\'un}, \citenamefont {Auffret}, \citenamefont
  {Vila}, \citenamefont {Gambarelli},\ and\ \citenamefont
  {Baltz}}]{Frangou2016}%
  \BibitemOpen
  \bibfield  {author} {\bibinfo {author} {\bibfnamefont {L.}~\bibnamefont
  {Frangou}}, \bibinfo {author} {\bibfnamefont {S.}~\bibnamefont {Oyarz\'un}},
  \bibinfo {author} {\bibfnamefont {S.}~\bibnamefont {Auffret}}, \bibinfo
  {author} {\bibfnamefont {L.}~\bibnamefont {Vila}}, \bibinfo {author}
  {\bibfnamefont {S.}~\bibnamefont {Gambarelli}}, \ and\ \bibinfo {author}
  {\bibfnamefont {V.}~\bibnamefont {Baltz}},\ }\href {\doibase
  10.1103/PhysRevLett.116.077203} {\bibfield  {journal} {\bibinfo  {journal}
  {Phys. Rev. Lett.}\ }\textbf {\bibinfo {volume} {116}},\ \bibinfo {pages}
  {077203} (\bibinfo {year} {2016})}\BibitemShut {NoStop}%
\bibitem [{\citenamefont {Lin}\ \emph {et~al.}(2016)\citenamefont {Lin},
  \citenamefont {Chen}, \citenamefont {Zhang},\ and\ \citenamefont
  {Chien}}]{Linweiwei2016}%
  \BibitemOpen
  \bibfield  {author} {\bibinfo {author} {\bibfnamefont {W.}~\bibnamefont
  {Lin}}, \bibinfo {author} {\bibfnamefont {K.}~\bibnamefont {Chen}}, \bibinfo
  {author} {\bibfnamefont {S.}~\bibnamefont {Zhang}}, \ and\ \bibinfo {author}
  {\bibfnamefont {C.~L.}\ \bibnamefont {Chien}},\ }\href {\doibase
  10.1103/PhysRevLett.116.186601} {\bibfield  {journal} {\bibinfo  {journal}
  {Phys. Rev. Lett.}\ }\textbf {\bibinfo {volume} {116}},\ \bibinfo {pages}
  {186601} (\bibinfo {year} {2016})}\BibitemShut {NoStop}%
\bibitem [{\citenamefont {Koon}(1997)}]{Koon_spinflop}%
  \BibitemOpen
  \bibfield  {author} {\bibinfo {author} {\bibfnamefont {N.~C.}\ \bibnamefont
  {Koon}},\ }\href {\doibase 10.1103/PhysRevLett.78.4865} {\bibfield  {journal}
  {\bibinfo  {journal} {Phys. Rev. Lett.}\ }\textbf {\bibinfo {volume} {78}},\
  \bibinfo {pages} {4865} (\bibinfo {year} {1997})}\BibitemShut {NoStop}%
\bibitem [{\citenamefont {Uchida}\ \emph {et~al.}(2015)\citenamefont {Uchida},
  \citenamefont {Qiu}, \citenamefont {Kikkawa}, \citenamefont {Iguchi},\ and\
  \citenamefont {Saitoh}}]{uchida_hight_T_SMR}%
  \BibitemOpen
  \bibfield  {author} {\bibinfo {author} {\bibfnamefont {K.}~\bibnamefont
  {Uchida}}, \bibinfo {author} {\bibfnamefont {Z.}~\bibnamefont {Qiu}},
  \bibinfo {author} {\bibfnamefont {T.}~\bibnamefont {Kikkawa}}, \bibinfo
  {author} {\bibfnamefont {R.}~\bibnamefont {Iguchi}}, \ and\ \bibinfo {author}
  {\bibfnamefont {E.}~\bibnamefont {Saitoh}},\ }\href
  {http://scitation.aip.org/content/aip/journal/apl/106/5/10.1063/1.4907546}
  {\bibfield  {journal} {\bibinfo  {journal} {Applied Physics Letters}\
  }\textbf {\bibinfo {volume} {106}},\ \bibinfo {eid} {052405} (\bibinfo {year}
  {2015})}\BibitemShut {NoStop}%
\bibitem [{\citenamefont {Zhou}\ \emph {et~al.}(2015)\citenamefont {Zhou},
  \citenamefont {Ma}, \citenamefont {Shi}, \citenamefont {Fan}, \citenamefont
  {Zheng}, \citenamefont {Evans},\ and\ \citenamefont {Zhou}}]{ZhouMPEAMR2015}%
  \BibitemOpen
  \bibfield  {author} {\bibinfo {author} {\bibfnamefont {X.}~\bibnamefont
  {Zhou}}, \bibinfo {author} {\bibfnamefont {L.}~\bibnamefont {Ma}}, \bibinfo
  {author} {\bibfnamefont {Z.}~\bibnamefont {Shi}}, \bibinfo {author}
  {\bibfnamefont {W.~J.}\ \bibnamefont {Fan}}, \bibinfo {author} {\bibfnamefont
  {J.-G.}\ \bibnamefont {Zheng}}, \bibinfo {author} {\bibfnamefont {R.~F.~L.}\
  \bibnamefont {Evans}}, \ and\ \bibinfo {author} {\bibfnamefont {S.~M.}\
  \bibnamefont {Zhou}},\ }\href@noop {} {\bibfield  {journal} {\bibinfo
  {journal} {Phys. Rev. B}\ }\textbf {\bibinfo {volume} {92}},\ \bibinfo
  {pages} {060402} (\bibinfo {year} {2015})}\BibitemShut {NoStop}%
\bibitem [{\citenamefont {V\'elez}\ \emph {et~al.}(2016)\citenamefont
  {V\'elez}, \citenamefont {Bedoya-Pinto}, \citenamefont {Yan}, \citenamefont
  {Hueso},\ and\ \citenamefont {Casanova}}]{Velez}%
  \BibitemOpen
  \bibfield  {author} {\bibinfo {author} {\bibfnamefont {S.}~\bibnamefont
  {V\'elez}}, \bibinfo {author} {\bibfnamefont {A.}~\bibnamefont
  {Bedoya-Pinto}}, \bibinfo {author} {\bibfnamefont {W.}~\bibnamefont {Yan}},
  \bibinfo {author} {\bibfnamefont {L.~E.}\ \bibnamefont {Hueso}}, \ and\
  \bibinfo {author} {\bibfnamefont {F.}~\bibnamefont {Casanova}},\ }\href@noop
  {} {\bibfield  {journal} {\bibinfo  {journal} {Phys. Rev. B}\ }\textbf
  {\bibinfo {volume} {94}},\ \bibinfo {pages} {174405} (\bibinfo {year}
  {2016})}\BibitemShut {NoStop}%
\bibitem [{\citenamefont {Lin}\ and\ \citenamefont {Chien}(2017)}]{wwlin2017}%
  \BibitemOpen
  \bibfield  {author} {\bibinfo {author} {\bibfnamefont {W.}~\bibnamefont
  {Lin}}\ and\ \bibinfo {author} {\bibfnamefont {C.~L.}\ \bibnamefont
  {Chien}},\ }\href@noop {} {\bibfield  {journal} {\bibinfo  {journal} {Phys.
  Rev. Lett.}\ }\textbf {\bibinfo {volume} {118}},\ \bibinfo {pages} {067202}
  (\bibinfo {year} {2017})}\BibitemShut {NoStop}%
\bibitem [{\citenamefont {Schulthess}\ and\ \citenamefont
  {Butler}(1998)}]{Schulthess1998}%
  \BibitemOpen
  \bibfield  {author} {\bibinfo {author} {\bibfnamefont {T.}~\bibnamefont
  {Schulthess}}\ and\ \bibinfo {author} {\bibfnamefont {W.}~\bibnamefont
  {Butler}},\ }\href {\doibase 10.1103/PhysRevLett.81.4516} {\bibfield
  {journal} {\bibinfo  {journal} {Physical Review Letters}\ }\textbf {\bibinfo
  {volume} {81}},\ \bibinfo {pages} {4516} (\bibinfo {year}
  {1998})}\BibitemShut {NoStop}%
\bibitem [{\citenamefont {Krug}\ \emph {et~al.}(2008)\citenamefont {Krug},
  \citenamefont {Hillebrecht}, \citenamefont {Haverkort}, \citenamefont
  {Tanaka}, \citenamefont {Tjeng}, \citenamefont {Gomonay}, \citenamefont
  {Fraile-Rodr\'{\i}guez}, \citenamefont {Nolting}, \citenamefont {Cramm},\
  and\ \citenamefont {Schneider}}]{Krug2008}%
  \BibitemOpen
  \bibfield  {author} {\bibinfo {author} {\bibfnamefont {I.~P.}\ \bibnamefont
  {Krug}}, \bibinfo {author} {\bibfnamefont {F.~U.}\ \bibnamefont
  {Hillebrecht}}, \bibinfo {author} {\bibfnamefont {M.~W.}\ \bibnamefont
  {Haverkort}}, \bibinfo {author} {\bibfnamefont {A.}~\bibnamefont {Tanaka}},
  \bibinfo {author} {\bibfnamefont {L.~H.}\ \bibnamefont {Tjeng}}, \bibinfo
  {author} {\bibfnamefont {H.}~\bibnamefont {Gomonay}}, \bibinfo {author}
  {\bibfnamefont {A.}~\bibnamefont {Fraile-Rodr\'{\i}guez}}, \bibinfo {author}
  {\bibfnamefont {F.}~\bibnamefont {Nolting}}, \bibinfo {author} {\bibfnamefont
  {S.}~\bibnamefont {Cramm}}, \ and\ \bibinfo {author} {\bibfnamefont {C.~M.}\
  \bibnamefont {Schneider}},\ }\href {\doibase 10.1103/PhysRevB.78.064427}
  {\bibfield  {journal} {\bibinfo  {journal} {Phys. Rev. B}\ }\textbf {\bibinfo
  {volume} {78}},\ \bibinfo {pages} {064427} (\bibinfo {year}
  {2008})}\BibitemShut {NoStop}%
\bibitem [{\citenamefont {Li}\ \emph {et~al.}(2014)\citenamefont {Li},
  \citenamefont {Tan}, \citenamefont {Ma}, \citenamefont {Yang}, \citenamefont
  {Arenholz}, \citenamefont {Hwang},\ and\ \citenamefont {Qiu}}]{LiJia2014}%
  \BibitemOpen
  \bibfield  {author} {\bibinfo {author} {\bibfnamefont {J.}~\bibnamefont
  {Li}}, \bibinfo {author} {\bibfnamefont {A.}~\bibnamefont {Tan}}, \bibinfo
  {author} {\bibfnamefont {S.}~\bibnamefont {Ma}}, \bibinfo {author}
  {\bibfnamefont {R.~F.}\ \bibnamefont {Yang}}, \bibinfo {author}
  {\bibfnamefont {E.}~\bibnamefont {Arenholz}}, \bibinfo {author}
  {\bibfnamefont {C.}~\bibnamefont {Hwang}}, \ and\ \bibinfo {author}
  {\bibfnamefont {Z.~Q.}\ \bibnamefont {Qiu}},\ }\href@noop {} {\bibfield
  {journal} {\bibinfo  {journal} {Phys. Rev. Lett.}\ }\textbf {\bibinfo
  {volume} {113}},\ \bibinfo {pages} {147207} (\bibinfo {year}
  {2014})}\BibitemShut {NoStop}%
\bibitem [{\citenamefont {Fraune}\ \emph {et~al.}(2000)\citenamefont {Fraune},
  \citenamefont {RÃŒdiger}, \citenamefont {GÃŒntherodt}, \citenamefont
  {Cardoso},\ and\ \citenamefont {Freitas}}]{NiOdomainsize}%
  \BibitemOpen
  \bibfield  {author} {\bibinfo {author} {\bibfnamefont {M.}~\bibnamefont
  {Fraune}}, \bibinfo {author} {\bibfnamefont {U.}~\bibnamefont {RÃŒdiger}},
  \bibinfo {author} {\bibfnamefont {G.}~\bibnamefont {GÃŒntherodt}}, \bibinfo
  {author} {\bibfnamefont {S.}~\bibnamefont {Cardoso}}, \ and\ \bibinfo
  {author} {\bibfnamefont {P.}~\bibnamefont {Freitas}},\ }\href@noop {}
  {\bibfield  {journal} {\bibinfo  {journal} {Applied Physics Letters}\
  }\textbf {\bibinfo {volume} {77}},\ \bibinfo {pages} {3815} (\bibinfo {year}
  {2000})}\BibitemShut {NoStop}%
\bibitem [{\citenamefont {Kim}\ \emph {et~al.}(2010)\citenamefont {Kim},
  \citenamefont {Jin}, \citenamefont {Wu}, \citenamefont {Park}, \citenamefont
  {Arenholz}, \citenamefont {Scholl}, \citenamefont {Hwang},\ and\
  \citenamefont {Qiu}}]{NiOFeXMLD}%
  \BibitemOpen
  \bibfield  {author} {\bibinfo {author} {\bibfnamefont {W.}~\bibnamefont
  {Kim}}, \bibinfo {author} {\bibfnamefont {E.}~\bibnamefont {Jin}}, \bibinfo
  {author} {\bibfnamefont {J.}~\bibnamefont {Wu}}, \bibinfo {author}
  {\bibfnamefont {J.}~\bibnamefont {Park}}, \bibinfo {author} {\bibfnamefont
  {E.}~\bibnamefont {Arenholz}}, \bibinfo {author} {\bibfnamefont
  {A.}~\bibnamefont {Scholl}}, \bibinfo {author} {\bibfnamefont
  {C.}~\bibnamefont {Hwang}}, \ and\ \bibinfo {author} {\bibfnamefont {Z.~Q.}\
  \bibnamefont {Qiu}},\ }\href {\doibase 10.1103/PhysRevB.81.174416} {\bibfield
   {journal} {\bibinfo  {journal} {Phys. Rev. B}\ }\textbf {\bibinfo {volume}
  {81}},\ \bibinfo {pages} {174416} (\bibinfo {year} {2010})}\BibitemShut
  {NoStop}%
\bibitem [{\citenamefont {Alders}\ \emph {et~al.}(1998)\citenamefont {Alders},
  \citenamefont {Tjeng}, \citenamefont {Voogt}, \citenamefont {Hibma},
  \citenamefont {Sawatzky}, \citenamefont {Chen}, \citenamefont {Vogel},
  \citenamefont {Sacchi},\ and\ \citenamefont {Iacobucci}}]{Alders1998}%
  \BibitemOpen
  \bibfield  {author} {\bibinfo {author} {\bibfnamefont {D.}~\bibnamefont
  {Alders}}, \bibinfo {author} {\bibfnamefont {L.}~\bibnamefont {Tjeng}},
  \bibinfo {author} {\bibfnamefont {F.}~\bibnamefont {Voogt}}, \bibinfo
  {author} {\bibfnamefont {T.}~\bibnamefont {Hibma}}, \bibinfo {author}
  {\bibfnamefont {G.}~\bibnamefont {Sawatzky}}, \bibinfo {author}
  {\bibfnamefont {C.}~\bibnamefont {Chen}}, \bibinfo {author} {\bibfnamefont
  {J.}~\bibnamefont {Vogel}}, \bibinfo {author} {\bibfnamefont
  {M.}~\bibnamefont {Sacchi}}, \ and\ \bibinfo {author} {\bibfnamefont
  {S.}~\bibnamefont {Iacobucci}},\ }\href {\doibase 10.1103/PhysRevB.57.11623}
  {\bibfield  {journal} {\bibinfo  {journal} {Physical Review B}\ }\textbf
  {\bibinfo {volume} {57}},\ \bibinfo {pages} {11623} (\bibinfo {year}
  {1998})}\BibitemShut {NoStop}%
\bibitem [{\citenamefont {Ganzhorn}\ \emph {et~al.}(2016)\citenamefont
  {Ganzhorn}, \citenamefont {Barker}, \citenamefont {Schlitz}, \citenamefont
  {Piot}, \citenamefont {Ollefs}, \citenamefont {Guillou}, \citenamefont
  {Wilhelm}, \citenamefont {Rogalev}, \citenamefont {Opel}, \citenamefont
  {Althammer}, \citenamefont {Gepr\"ags}, \citenamefont {Huebl}, \citenamefont
  {Gross}, \citenamefont {Bauer},\ and\ \citenamefont
  {Goennenwein}}]{Granzhorn2016}%
  \BibitemOpen
  \bibfield  {author} {\bibinfo {author} {\bibfnamefont {K.}~\bibnamefont
  {Ganzhorn}}, \bibinfo {author} {\bibfnamefont {J.}~\bibnamefont {Barker}},
  \bibinfo {author} {\bibfnamefont {R.}~\bibnamefont {Schlitz}}, \bibinfo
  {author} {\bibfnamefont {B.~A.}\ \bibnamefont {Piot}}, \bibinfo {author}
  {\bibfnamefont {K.}~\bibnamefont {Ollefs}}, \bibinfo {author} {\bibfnamefont
  {F.}~\bibnamefont {Guillou}}, \bibinfo {author} {\bibfnamefont
  {F.}~\bibnamefont {Wilhelm}}, \bibinfo {author} {\bibfnamefont
  {A.}~\bibnamefont {Rogalev}}, \bibinfo {author} {\bibfnamefont
  {M.}~\bibnamefont {Opel}}, \bibinfo {author} {\bibfnamefont {M.}~\bibnamefont
  {Althammer}}, \bibinfo {author} {\bibfnamefont {S.}~\bibnamefont
  {Gepr\"ags}}, \bibinfo {author} {\bibfnamefont {H.}~\bibnamefont {Huebl}},
  \bibinfo {author} {\bibfnamefont {R.}~\bibnamefont {Gross}}, \bibinfo
  {author} {\bibfnamefont {G.~E.~W.}\ \bibnamefont {Bauer}}, \ and\ \bibinfo
  {author} {\bibfnamefont {S.~T.~B.}\ \bibnamefont {Goennenwein}},\ }\href@noop
  {} {\bibfield  {journal} {\bibinfo  {journal} {Phys. Rev. B}\ }\textbf
  {\bibinfo {volume} {94}},\ \bibinfo {pages} {094401} (\bibinfo {year}
  {2016})}\BibitemShut {NoStop}%
\bibitem [{\citenamefont {Takei}\ \emph {et~al.}(2014)\citenamefont {Takei},
  \citenamefont {Halperin}, \citenamefont {Yacoby},\ and\ \citenamefont
  {Tserkovnyak}}]{takeiPRB14}%
  \BibitemOpen
  \bibfield  {author} {\bibinfo {author} {\bibfnamefont {S.}~\bibnamefont
  {Takei}}, \bibinfo {author} {\bibfnamefont {B.~I.}\ \bibnamefont {Halperin}},
  \bibinfo {author} {\bibfnamefont {A.}~\bibnamefont {Yacoby}}, \ and\ \bibinfo
  {author} {\bibfnamefont {Y.}~\bibnamefont {Tserkovnyak}},\ }\href {\doibase
  10.1103/PhysRevB.90.094408} {\bibfield  {journal} {\bibinfo  {journal} {Phys.
  Rev. B}\ }\textbf {\bibinfo {volume} {90}},\ \bibinfo {pages} {094408}
  (\bibinfo {year} {2014})}\BibitemShut {NoStop}%
\bibitem [{\citenamefont {Chen}\ \emph {et~al.}(2016)\citenamefont {Chen},
  \citenamefont {Takahashi}, \citenamefont {Nakayama}, \citenamefont
  {Althammer}, \citenamefont {Goennenwein}, \citenamefont {Saitoh},\ and\
  \citenamefont {Bauer}}]{chenJPCM16}%
  \BibitemOpen
  \bibfield  {author} {\bibinfo {author} {\bibfnamefont {Y.-T.}\ \bibnamefont
  {Chen}}, \bibinfo {author} {\bibfnamefont {S.}~\bibnamefont {Takahashi}},
  \bibinfo {author} {\bibfnamefont {H.}~\bibnamefont {Nakayama}}, \bibinfo
  {author} {\bibfnamefont {M.}~\bibnamefont {Althammer}}, \bibinfo {author}
  {\bibfnamefont {S.~T.~B.}\ \bibnamefont {Goennenwein}}, \bibinfo {author}
  {\bibfnamefont {E.}~\bibnamefont {Saitoh}}, \ and\ \bibinfo {author}
  {\bibfnamefont {G.~E.~W.}\ \bibnamefont {Bauer}},\ }\href
  {http://stacks.iop.org/0953-8984/28/i=10/a=103004} {\bibfield  {journal}
  {\bibinfo  {journal} {Journal of Physics: Condensed Matter}\ }\textbf
  {\bibinfo {volume} {28}},\ \bibinfo {pages} {103004} (\bibinfo {year}
  {2016})}\BibitemShut {NoStop}%
\bibitem [{\citenamefont {Hung}\ \emph {et~al.}(2017)\citenamefont {Hung},
  \citenamefont {Hahn}, \citenamefont {Chang}, \citenamefont {Wu},
  \citenamefont {Ohldag},\ and\ \citenamefont {Kent}}]{SMR_RT_Kent}%
  \BibitemOpen
  \bibfield  {author} {\bibinfo {author} {\bibfnamefont {Y.-M.}\ \bibnamefont
  {Hung}}, \bibinfo {author} {\bibfnamefont {C.}~\bibnamefont {Hahn}}, \bibinfo
  {author} {\bibfnamefont {H.}~\bibnamefont {Chang}}, \bibinfo {author}
  {\bibfnamefont {M.}~\bibnamefont {Wu}}, \bibinfo {author} {\bibfnamefont
  {H.}~\bibnamefont {Ohldag}}, \ and\ \bibinfo {author} {\bibfnamefont {A.~D.}\
  \bibnamefont {Kent}},\ }\href@noop {} {\bibfield  {journal} {\bibinfo
  {journal} {AIP Advances}\ }\textbf {\bibinfo {volume} {7}},\ \bibinfo {pages}
  {055903} (\bibinfo {year} {2017})}\BibitemShut {NoStop}%
\end{thebibliography}

%


\end{document}